# Sneak into Devil's Colony-Study of Fake Profiles in Online Social Networks and the Cyber Law

Mudasir Ahmad Wani [a] mudasirwanijmi@gmail.com

Suraiya Jabin [b] sjabin@jmi.ac.in

Ghulam Yazdani [c] gyazdani@jmi.ac.in

Nehaluddin Ahmad [d] profnehaluddin.ahmad@unissa.edu.bn

[a, b] **Department of Computer Science**
Jamia Millia Islamia (Central University), New Delhi, India

[c] **Faculty of Law**
Jamia Millia Islamia (Central University), New Delhi, India

[d] **Faculty of Law and Sharia**
Sultan Sharif Ali Islamic University, (UNISSA) Brunei Darussalam

***Abstract***

Massive content about user's social, personal and professional life stored on Online Social Networks (OSNs) has attracted not only the attention of researchers and social analysts but also the cyber criminals. These cyber criminals penetrate illegally into an OSN by establishing fake profiles or by designing bots and exploit the vulnerabilities of an OSN to carry out illegal activities. With the growth of technology cyber crimes have been increasing manifold. Daily reports of the security and privacy threats in the OSNs demand not only the intelligent automated detection systems that can identify and alleviate fake profiles in real time but also the reinforcement of the security and privacy laws to curtail the cyber crime.

In this paper, we have studied various categories of fake profiles like compromised profiles, cloned profiles and online bots (spam-bots, social-bots, like-bots and influential-bots) on different OSN sites along with existing cyber laws to mitigate their threats. In order to design fake profile detection systems, we have highlighted different category of fake profile features which are capable to distinguish different kinds of fake entities from real ones. Another major challenges faced by researchers while building the fake profile detection systems is the unavailability of data specific to fake users. The paper addresses this challenge by providing extremely obliging data collection techniques along with some existing data sources. Furthermore, an attempt is made to present several machine learning techniques employed to design different fake profile detection systems.

## 1. INTRODUCTION

Online Social Network Analysis (OSNA) is considered as one of the most emerging research fields. An Online Social Network (OSN) is the grouping of nodes (individuals, actors, organizations, nations, states or web pages etc.) around the world connected by a set of links (relationships, interactions, distances, hyperlinks, etc). Basically, OSNs refer to the web applications which are primarily designed to facilitate interaction, collaboration and share content among users. OSNs have changed the way people think, express, and socialize with outside world. Nowadays there are a huge number of social networking sites like Facebook, Twitter, Flicker, LinkedIn, Researchgate, etc. which are used by people to carry out their social and professional activities. Since the structure of OSNs bears a resemblance to the real-life communities and they hold a massive amount of user content, therefore, they are highly important to the researchers and several other disciplines including marketing, sociology, politics, etc. Marketing companies study OSNs to design viral marketing strategies and attain their potential customers, sociologists use them to analyze the human behavior and politicians use them to empower their political campaigns. [95,100,101].

Apart from researchers and business organizations, the universal popularity of OSNs has also attracted the attention of social criminals. These social criminals (or simply cyber criminals) exploit the exposure and weakness of an OSN to perform unlawful, misleading, malicious, or discriminatory operations. They penetrate into the social network either by creating fake profiles or by executing a number of identity theft attacks like cloning attacks, spoofing attacks [24], etc. on existing users to steal their credentials. These cybercriminals especially the professional attackers also design different types of bots to control their fake profiles without much human effort [78]. There are various kinds of fake profiles, and their functions generally vary with the type of network on which they have been established. A growing number of hackers are creating forged identities on networks like Facebook and Twitter in order to access the social as well as personal information of users, to promote a particular brand or a person, to defame a user, etc. Adversaries may target professional sites like LinkedIn and Researchgate with the aim to track the activity of the members or to gain the trust of business professionals, lure users to give up personal details1. They also target dating websites with the intend to develop personal, romantic, or sexual relationships or gain financial benefits, gifts, or personal information, etc. A study [114] presented a review of the different security and privacy risks to OSN users, and endow with a simple set of recommendations to safeguard user virtual as well as real world.

Although, fake profiles are not always harmful; users sometimes create additional profiles for fun and entertainment, or for connecting with specific friend group only, etc. But as they violate the rules and regulations of the service, they are considered as illegal. Here rules and regulations in the context of OSN may mean the owner should not have more than one personal account, it should not spread any unlawful or malicious content, it should not collect the user's information or access network by automated means such as bots and spiders [10, 76], etc. According to Facebook an account maintaining by a person other than his principal account is fake[2]**.** Millions of fake profiles exist on popular social networks like Facebook and Twitter especially in the markets of China and India[3]. Social networking service providers are employing a number of ways to ensure the user security. For example, Facebook has provided several options to enhance the privacy of user accounts like protecting the password and sending location-specific login alerts and location alerts. Users can also use the extra security features of the network like how to log- out from remote devices, how to keep Facebook password safe using app passwords, etc. [87]. Besides that, Facebook also has its own inbuilt immune system [27] to detect objectionable profiles on the network. Similarly, Twitter [82] and LinkedIn [83] also allow their users to report a recognized spam or a fugitive content. Recently Facebook introduced an Artificial Intelligence-based system called Deep Text Tool [84] which is able to understand the text like humans. Besides helping the users with what they want to say, the tool would also be able to help in filtering the spam content in near future. Furthermore, Instagram is developing an anti-harassment tool [86] to filter comments or even help users to turn off the comment option on a particular post. This way the users can block a particular comment to avoid harassment. A study proposed a framework called "Safebook"[85] to protect user's personal data from both the malicious users as well as service providers who violate privacy rules.

Apart from an enormous number of methods for the detection of malicious users, there are plenty of cyber laws for cybercriminals. The cyber laws are usually framed by government IT and cyber experts in every country. As per ministry of electronics and information technology, government of India, destroying or altering the computer network, computer program or computer source code is punishable with imprisonment up to three years, or with fine which may extend up to two lakh rupees, or with both [106]. Also, as per the section 66F of the Information Technology Act 2000[4], whoever commits or conspires to commit cyber terrorism shall be punishable with imprisonment which may extend to imprisonment for life. Similarly, we have a cyber law for identity theft, under section 66C of IT Act, 2000[5], which states that the person shall be punished with imprisonment of either description for a term which may extend to three years and shall also be liable to fine which may extend to rupees one lakh (INR), whoever, fraudulently or dishonestly make use of the electronic signature, password or any other unique identification feature of any other person. Since there are very strict regulations and punishments for different category of cyber criminals but still this cyber crime especially cyber terrorism exists in every nation. The Information Technology Act in India has proved to be inadequate to a certain extent during its application. Adversaries are easily hacking into banking systems, social networking websites, e-commerce websites, etc. [107]. The tools and instruments needed to investigate cybercrime are quite different from those used to investigate ordinary crimes.

---

1 http://www.bbc.com/news/technology-34994858

2 https://www.facebook.com/legal/terms

3 http://www.gadgetsnow.com/tech-news/Facebook-may-have-over-100-million-fake-accounts-globally/articleshow/34672084.cms

4 http://www.itlaw.in/section-66f-punishment-for-cyber-terrorism/

5 https://cybercrimelawyer.wordpress.com/category/66c-punishment-for-identity-theft/

In order to control the cyber terrorism, more advanced automated methods for fake profile detection are needed. Also, the cyber laws need to be strengthened to handle cybercrimes across the boundaries. Cyber crime is a global phenomenon and therefore it should be tackled on the same level. Collectively it has been observed that the need of the hour is to identify a unique set of features to design effective model for detection of fake profiles on social networks and a worldwide uniform cyber laws in order to combat cyber crimes. In this paper, we aim to put everything about online malicious accounts at one place along with different cyber laws and commandments especially in Indian jurisdiction to curtail the fake profiles and their cybercrime.

OSNs are playing a vital role in contemporary life, people relying on them from different dimensions including social interactions, information sharing, and other daily activities. The negative impact on these OSNs by Fake profiles not only damages the user experience, but also the marketability and advertising potential of the given OSN. In respect to above circumstances, following are some key aspects which motivate this work to survey different kinds of suspicious identities on OSNS.

- Online Fake profiles exist on every social networking platform with diverse aims. The different nature of fake profiles and the way they achieve their goals is still at its infancy. Furthermore, the cyber laws against cyber criminals are still generalised, inadequate with several shortcomings such as, there is ambiguity in terms, clear definitions for the important terms in law are not mentioned which can be dangerous and may have several degrees of interpretation.

- As online fake users carry out their illicit operations based on their aim and architecture of the platform (online social networking websites), therefore, they may cuddle an exclusive set of attributes. Study of unique characteristics about every category of fake profiles can better aid in building efficient detection systems.

- One of the major hurdles faced by researchers while building the fake profile detection systems is the unavailability of data specific to fake users on OSNs. Although very few, but there exist obliging data collection techniques which should be available to researchers in order to obtain the data to learn the fake profile detection models.

- A number of studies towards detection of suspicious identities have been carried out so far using different machine learning based techniques on different platforms. The evolving research in this domain demands the appropriate techniques to deal with all kinds of bogus users. Therefore, all the techniques along with types of fake profiles should be available at a single place which assists to solve a particular problem with appropriate set of techniques.

Based on the above aspects the aim of this paper is to collect everything allied to online phony users at a single place. The rest of the paper is structured as follows: Section 2 describes the various categories of fake profiles found on different social networking sites along with existing cyber laws to alleviate the threats caused by them. Section 3 presents different types of features employed by researchers so far for the detection of fake profiles on OSNs. Several data collection methods specific to fake profiles have been discussed in section 4. Section 5 explains different fake profile detection techniques. Finally, section 6 concludes the paper.

## Online Social Network (OSN) Fake Profiles

To avail the service, most of the OSNs require a user to establish a profile on the network containing his/her basic (sometimes personal) information such as name, gender, location, e-mail address, etc. The openness of these social networking sites enables adversaries to exploit the service by creating different kinds of fake profiles to carry out illegitimate, adversarial, unlawful, misleading or malicious activities such as spamming, promotion and advertising, stalking, bullying, defaming, etc. However, specific reasons behind establishing the fake profiles generally depend on the type of the social network being targeted. Adversary creates forged identities on networks like Facebook and Twitter to access the personal information of users, endorse a particular brand or a person or to defame a user, etc. For professional sites like LinkedIn and Researchgate, they aim to track the activity of the members or to gain the trust of business professionals. Attackers often target dating websites to take the advantage of people who are seeking for ideal matches and work collaborators to extract money from these users by playing with their emotions or stealing personal information. One of the most dangerous fake profiles on dating OSNs is called Catfisher [108], a person who uses the online dating websites to tempt people into a scam romance.

The Indian cyber law stated specific provisions for different unlawful operations executed by means of fake profiles. According to the section 449 and 500, Indian Panel Code (IPC)[6], whoever, by words either spoken or intended to be read, or by signs or by visible representations, makes or publishes any imputation concerning any person intending to harm, or knowing or having reason to believe that such imputation will harm, the reputation of such person, is said, except in the cases hereinafter excepted to defame that person will be punishable with imprisonment of two years or fine or both. Similarly, the section 469, Indian Panel Code (IPC)[7] deals with the forgery accounts and states that if anyone creates a false document or fake account by which it harms the reputation of a person. The punishment of this offense can extend up to 3 years and fine. Furthermore, making a random fake account is not punishable unless it is fraudulent under the cyber laws of the country. Under section 66A (b) of Information Technology act 2000, a person is considered as guilty of sending information which he knows to be bogus and is aimed to cause annoyance, inconvenience, danger etc. The punishment for such activities is imprisonment up to three years and a fine could also be levied.

This section provides an exhaustive classification of a different variety of fake profiles with an emphasis on online social networks, along with cyber laws to curb their consequence.

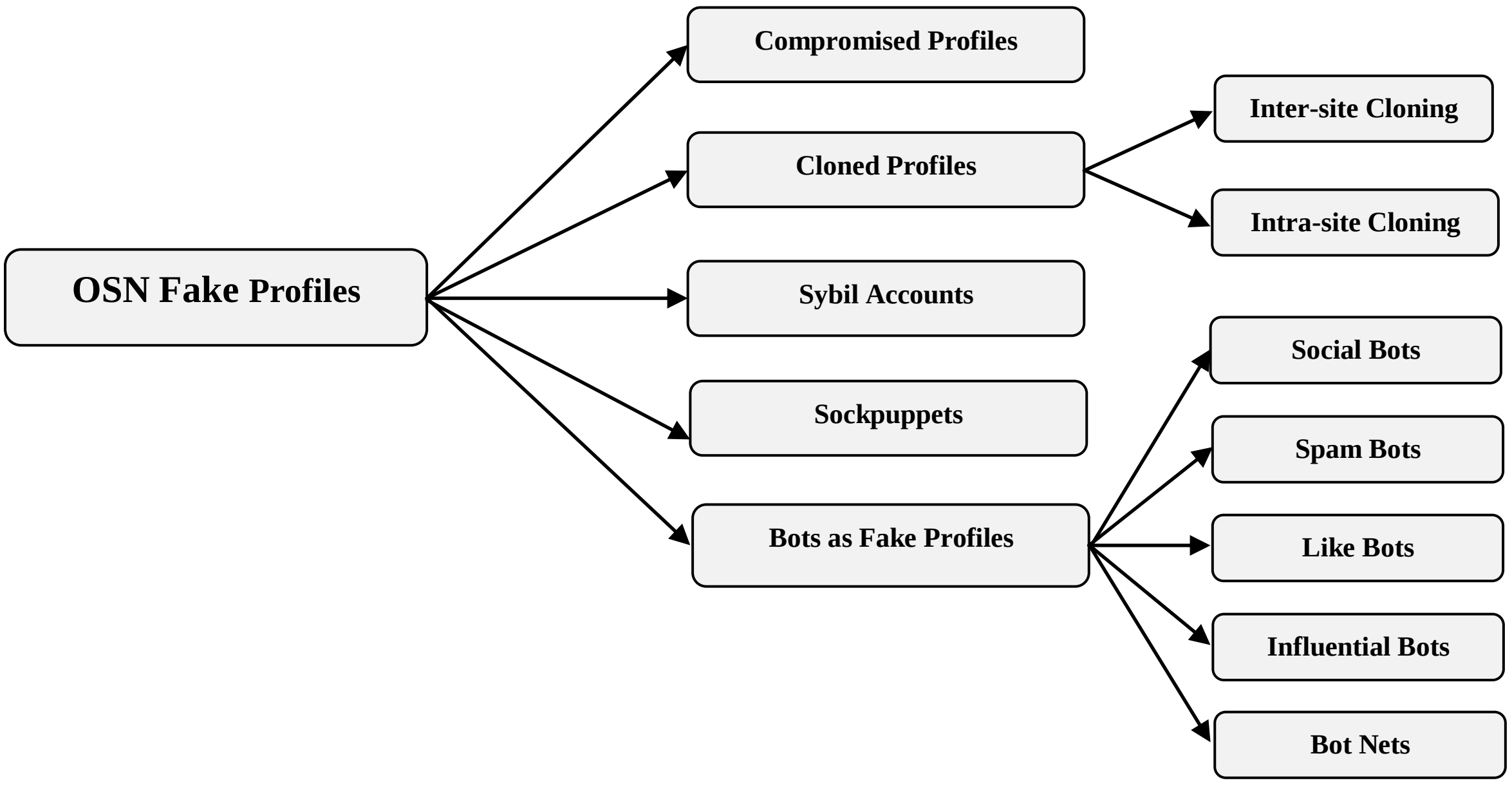

**Figure 1:** Types of fake profiles in online social networks

As shown in Figure 1, the fake profiles have been divided into five categories viz compromised profiles, cloned profiles, sybil accounts, sockpuppets, and fake bot profiles which are briefly described in following subsections.

[6] http://bombayhighcourt.nic.in/libweb/oldlegislation/ipc1860/Chapter%2021.pdf
[7] http://www.indianlawcases.com/Act-Indian.Penal.Code,1860-1929

These categories can be considered as the different ways by which the adversaries achieve their ill aims on different online social networking platforms.

### 2.1 Compromised Profiles

Compromised accounts are basically the real accounts but their owners don't have complete control over them and they have lost the control to a phisher or any malware agent [7]. As per the Facebook terms and conditions, any legitimate account that is accessed by the person who is not the authorized owner of the account is considered as compromised. According to authors in [58], compromised accounts are the most difficult type of accounts to be detected as the real owner has already maintained a level of trust on the networks. Another recent study [23] states that more than 97% profiles are compromised rather than fake which are generally used to spread spam. The fake profiles are generally created to steal the credentials from the real users, and then these fake profiles are abandoned or deactivated and compromised ones are used by adversaries.

Compromised profiles have much value because they have already established a level of trust within their network and therefore, they cannot be easily detected and alleviated from the social network by the service providers. The study [23] reveals that the compromised real profiles spread more malicious content than other types of fake profiles. Facebook assists its users to recover hacked and compromised accounts once reported. There are options such as my-account-was-hacked[8] or *My-Account-is-Compromised*[9] on the Facebook help page, using which the users can report their compromised accounts.

Usually, the cyber criminals launch various phishing attacks to obtain credentials of a real account in order to perform several unlawful activities, and this is considered as a serious cybercrime. According to the section 66D of the Information Technology Act, 2000 whoever, by means of any communication device or computer resource cheats by impersonating, shall be punished with imprisonment which may extend to three years and shall also be liable to fine which may extend to one lakh rupees in Indian currency. Furthermore, section 66C of Information Technology Act 2000, (amendment 2008) states that if any person, fraudulently or dishonestly uses password or any other unique identification feature belonging to any other person, shall be punished with imprisonment of either description for a term which may extend to three years and shall also be liable to fine which may extend to amount rupees one million in Indian currency.

There are several reasons for a profile to get compromised such as weak passwords, virus infections, sharing passwords, etc. Users should take proper care while using the social media accounts to secure their personal and social data from cybercriminals.

### 2.2 Cloned Profiles

Profile cloning is a technique in which the adversary establishes another profile using the information such as name, age, gender, profile picture, etc. of any existing real profile. In other words, we can say profile cloning is the process of stealing the victim's information in order to create one more profile to spread spam, obtain private information about the victim and victim's friends or carry out other scams including, stalking, defaming, etc. Once the clone profile is created, the cloner can send friend requests to people in the victim's friends list and start sending scam messages on the name of the victim. A clever cloner may even perform identity theft by tricking the victim's friends into revealing a large amount of their personal and financial information. This is called as Identity Clone Attacks (ICAs) [24]. There are two types of profile cloning attacks namely single site profile cloning and cross-site profile cloning. The attackers are usually well funded, skilled persons and have almost everything available at their disposal and have control over compromised and infected accounts [27]. The adversary can be a strange person, but statistics show that adversary has the knowledge of victim and can be one of the victim's relatives, friend or colleague [24]. The two types of cloning attacks (intra site and inter-site profile cloning) are shown in Figure 2 and Figure 3, respectively, and are briefly explained in the following subsections.

[8] https://www.facebook.com/help/131719720300233
[9] https://www.facebook.com/hacked

### 2.2.1 Intra site profile cloning:

In case of intra site cloning, the adversary creates one more profile of the victim on the same network and sends the friend requests or follows the victim's friends. The victim's friends easily accept the friend request by treating it as a request from a legitimate user. It is possible to have different online accounts with the same name because in real life multiple persons can have the same name with different contact details, mobile number, and email address. Adversaries are taking the advantage by creating one more account of an already existing person and pretending to be some real person with the same name.

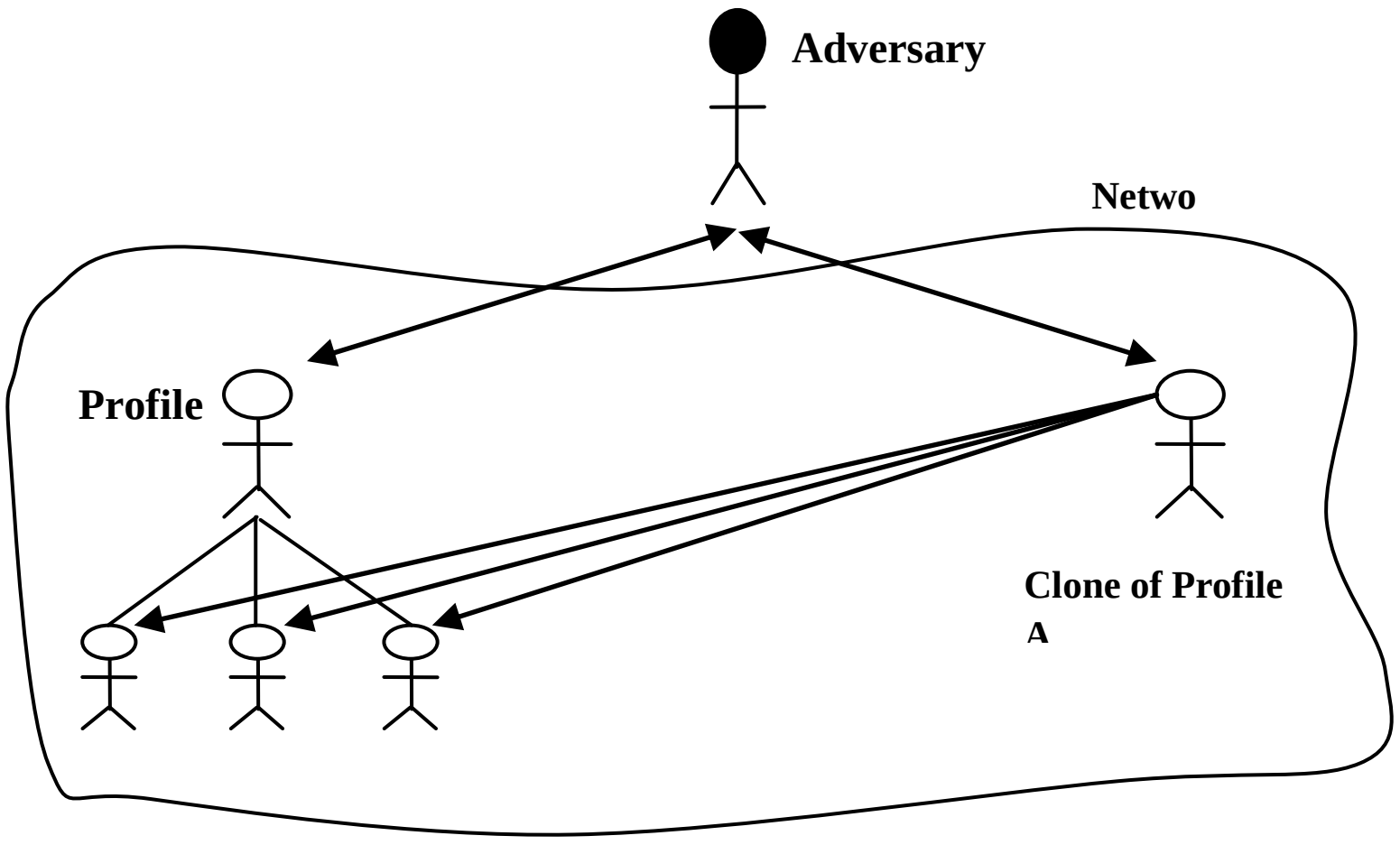

**Figure 2: Intra site or same site profile cloning**

### 2.2.2 Inter site profile cloning:

The inter-site cloning (also called cross-site cloning) is the process in which the adversary creates one more profile of victim on the new network where the user is not yet registered and sends the request to the victim's friends who are on both the networks. Inter site profile cloning can also be viewed as the reconstruction of victim's friend list in another social network where he/she is yet not registered. The main goal of the adversary in creating these cloned profiles is to steal people's personal information, deceive or defame others or sometimes simply for entertainment. These ICAs are a matter of concern for both users as well as service providers as it becomes very difficult to detect these kinds of attacks. The users simply treat it as a friend request from a legitimate user and the service providers take it as a new user registering in these online social networks [6]. Detection of cloned profiles with more accuracy can enhance the level of security in OSNs which in turn will keep users safe from any kind of unauthorized access.

A recent study [28] has suggested different ways to cope with cloning attacks and recommended measures for OSN sites to improve the security and how the users can protect themselves. Bilge et al.[6] presented two automated ICAs namely 'profile cloning' and 'cross-site profile cloning' and proposed prototype attack system (iCLONER) to attack the five most popular OSNs including XING, StudVZ, MeinVZ, Facebook and LinkedIn. This study showed that ICA schemes are much effective and enemies do not raise much suspicion in users.

Profile cloning is a serious issue in online social networks. Normal users are not aware that their identities are being copied and used as a weapon to destroy their own kingdom by dodgy characters.

These criminals actually copy all the content from victim's profile including profile name, profile picture, education, work even status updates to give it exactly the same look as the real account and exploits it to perform other cyber crimes.

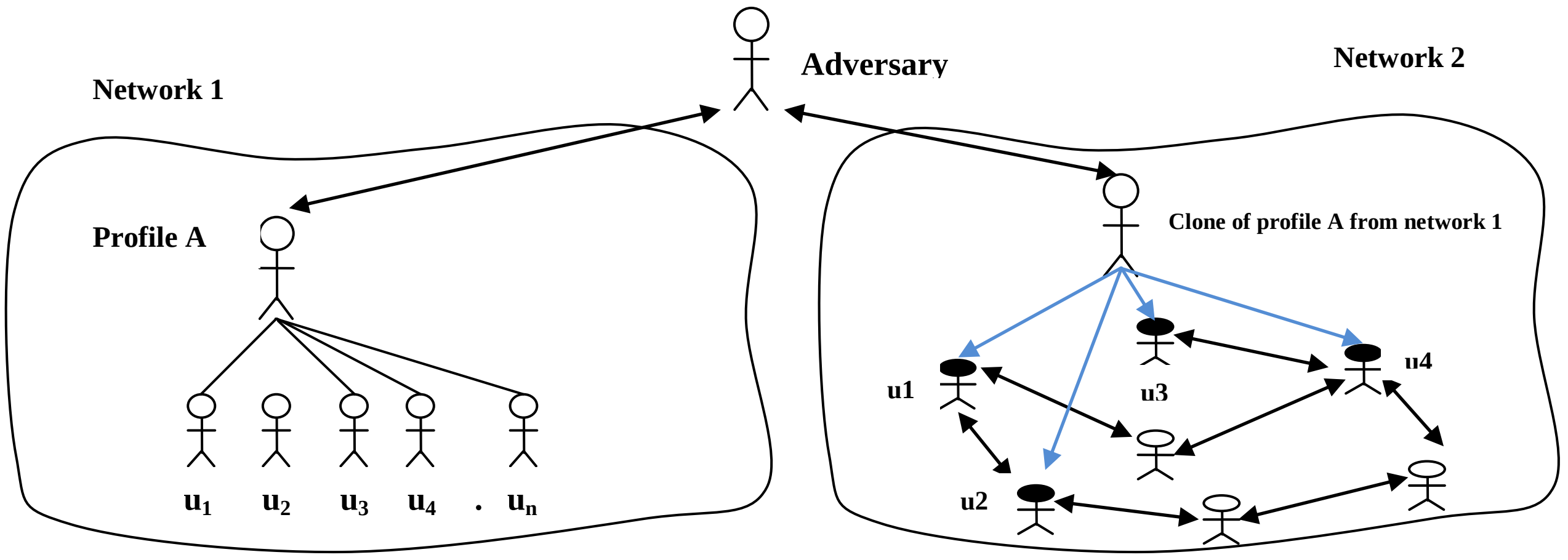

**Figure 3: Inter site or Cross site Profile Cloning**

As per section 66C of Information Technology Act 2000, (amendment 2008), if any person, fraudulently or dishonestly uses electronic signature, password or any other unique identification feature belonging to any other person, shall be punished with imprisonment of either description for a term which may extend to three years and shall also be liable to fine which may extend to amount rupees one million in Indian currency.

There are various approaches and techniques [6][28][109] for the detection of cloned profiles in online social networks but still, the profiles are cloned and misused at very higher rates. The behavior of these accounts needs to be studied more rigorously to spot and stop profile cloning.

### 2.3 Sockpuppets:

A sockpuppet is an account developed with an aim to deceive others or to promote someone or something on discussion forums, blogs, social networking sites etc. In other words, sockpuppets are those online accounts which are created to cheat the netizens in different ways, for example, to make the people believe that a particular product is good to buy, there is low risk in an investment plan with high return[40] etc. Usually, in case of OSN sites, the blocked users create new accounts which are referred as sockpuppetry [2]. According to the authors in [40], if there exist two different accounts on any news blog, social network or any discussion forum that belong to the same person, it is called sockpuppet pair. Sockpuppets are created for several reasons including business promotions, fake reviews on books and movies, false campaigning, defend or support a person or an organization, etc. In case of discussion forums, sockpuppets are used to engage people by deceiving others or manipulating discussions. The authors in [104] studied the sockpuppetry and showed that sockpuppets have different posting behavior than normal users in various discussion forums. In OSNs like Facebook, Twitter, etc., Sockpuppets are created for the purpose of making more followers, false likes and also for the purpose of conducting mass propaganda through retweets and comments. Establishing Sockpuppets is considered as one of the main ways of online deception [105]. Nowadays, sockpuppets are used for false marketing- an example of astroturfing[10], in which the people artificially stimulate online conversation and positive reviews about a particular product, brand or service. Because sock puppets can be created quickly and do not need manual maintenance, they are often used on social media sites to attract public interest or defame a competitor's product, brand or service.

Astroturfing and sockpuppetry, in general, are unethical and illegal. If detected, sockpuppet marketing can have a negative impact, causing potential customers to lose trust and doubt if the product or service is so lacking in value that it cannot be effectively promoted honestly. In the United States, the Federal Trade Commission (FTC) under Section 5[102] has the legal authority to charge fines if a company found involved in sockpuppet marketing.

### 2.4 Sybil Accounts

[10] https://www.merriam-webster.com/dictionary/astroturfing

In case of sybil accounts, the malicious users create multiple accounts and handle them manually to attack the trusted network. When a node in online social network claims multiple roles and threatens the security, it is referred as Sybil attack. In a Sybil attack, the attacker weakens the reputation of a network by manually establishing and maintaining a large number of pseudonymous identities, using them to spread the malware and spam on social networks and to gain a disproportionately large influence. Sybil attackers have many goals like bad mouthing an opinion, illegal voting, accessing resources, compromising the security and privacy, etc [36]. According to [37], social networks with well-defined community structure are more exposed to these Sybil attacks because their links can be used by the Sybil attackers more effectively. Several studies [36, 37, 38 42, 68,77] have been carried out so far for the defense of these attacks, but still, the detection of Sybil attacks is in its early stage.

Disrupting or deceiving the reputation or key functions of a system, network or an individual is considered as offense according to the section 415[11], Indian panel Code (IPC). According to this section whoever, by deceiving any person, fraudulently or dishonestly induces the person so deceived to the extent to deliver any property to any person, or to consent that any person shall retain any property, or intentionally induces the person so deceived to do or omit to do anything which he would not do or omit if he were not so, and which act or omission causes or is likely to cause damage or harm to that person in body, mind, reputation or property, is said to "cheat" and under the section 417[12] Indian panel Code (IPC), whoever cheats shall be punished with imprisonment of either description for a term which may extend to one year, or with fine, or with both.

### 2.5 Bots as Fake Profiles

A bot is a computer program that runs different scripts to perform human activities over the internet. According to the authors in [46], a bot is a computer program that produces some data to interact with humans especially the persons using the internet (netizens) in order to alter their behavior. The main use of bots is web data crawling where a simple online computer program identifies and extracts the information from web servers at a much higher speed which was not possible by a human alone. More than 60% of the total web data is generated by bots [78]. But nowadays the bots have been exploited by spammers on different social networks to execute various malicious activities and turned to be a serious threat to the internet. According to a study [10], more than 8% bots exist in the Twitter network and most of them have been developed for commercial purposes. The cyber criminals establish fake profiles on OSNS and control them by an automated program for performing several malicious activities. Bot profiles are used to retweet a post without verifying its source in order to make it viral. In online multiplayer games, bots are used to gain the unfair advantage [44, 46]. Sometimes bots act as automated avatars to interact with humans and create social networks, which are even more difficult to identify [45]. Bots can also be used to influence users, post messages and send friend requests [47] in online social networks. From the working point of view, bots are similar as Sybil accounts but the main difference is Sybil accounts are handled by users manually whereas bots are automated computer programs [46]. Many researchers are working on the detection of bots in order to mitigate their bad effect. The authors in [99] have deeply studied the behavior of bot-controlled Twitter accounts and highlighted how bots use different retweet and mentioned strategies while interacting with humans or other bots on the network and presented a framework to detect such accounts.

Various OSN service providers employed a number of ways to fight the spam bots. Facebook has its Facebook Immune System (FIS) [27] to deal with bots. However, the users in various OSNs claim that their legitimate accounts are being caught by the detection techniques.

It is not true that a bot is always designed for malicious activities. They can be used to assist the internet users as well. For example, chatbots [103] can be used to help students to answer their day-to-day queries and the bots which are developed for daily activities like weather update (e.g. Twitter bots) are the examples of good bots. But unfortunately, cyber criminals exploit the functionalities of these bots to use them as fake profiles in order to perform various unlawful, misleading, malicious operations. Based on the functionality we present five categories of fake bot profiles as shown in Figure 4. Each category is discussed in the following subsections.

[11] http://thepracticeoflawjalan.blogspot.in/2012/04/offences-cheating.html
[12] http://devgan.in/ipc/section/417/

### 2.5.1 Spam Bots

Spam bot is a computer program specially designed to spread malicious content such as links to personal blogs, paid contents, pornographic websites, and advertisements, or to shill for any person or organization, etc. by creating a large number of unwanted relationships on the network [50].

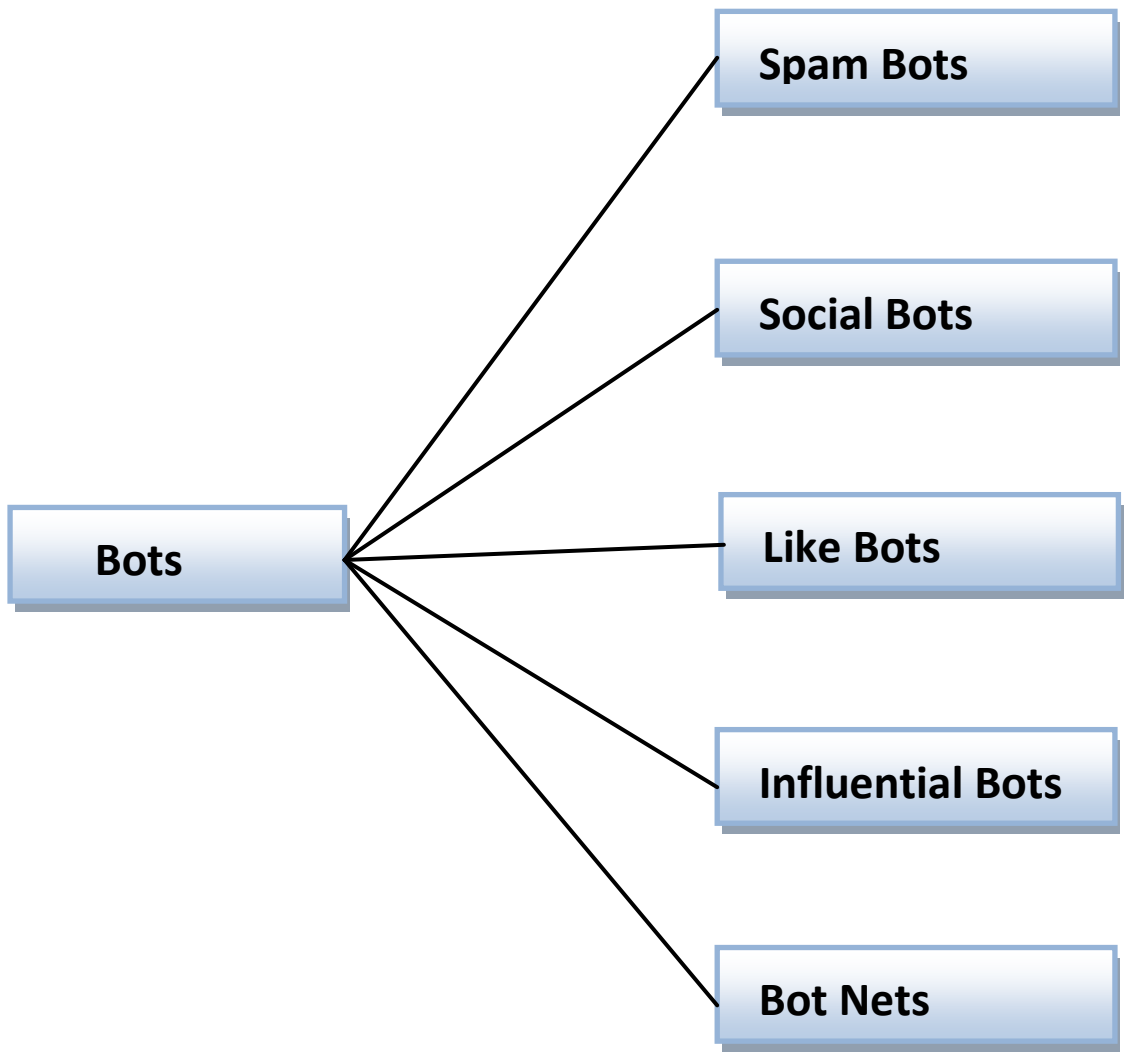

**Figure 4: Categories of Bots in OSNs**

In the [44] the authors studied the behavior of spam bots in the Twitter network and applied several classification techniques to differentiate them from normal bots. One more study in [46] has presented a method known as *BotOrNot* to differentiate between a human and a bot controlled Twitter account based on six categories of features viz Network, User, Friends, Temporal, Content and Sentiment features. Although the spam bots are new to the OSNs, the detection of spams has been previously focused on the emails, web sites, etc.

Initially, spam-bots were designed to gather or harvest, e-mail addresses from the Internet for sending unsolicited e-mail, also known as spam. According to the CAN-SPAM Act of 2003[98], the Federal Trade Commission (FTC) has the authority to levy fines up to $11,000 against business owners engaging in commercial emails. Also according to Information Technology Act, section 43(b), section 66 read with section 43(d), (e), (f)[13], sending of spam dishonestly or fraudulently is punishable with “imprisonment up to three years or fine up to five lakh rupees or both”

### 2.5.2 Social Bots

Social bots are the computer programs used by humans for their several online activities. According to a study [49], social bots are highly complex computer programs which behave like humans and usually keep users busy. Social bots are the programs which publicize themselves like viruses to reach and infect a maximum number of users [47]. One more study [50] refers to them as bots which control accounts on online social networks and imitate the behavior of legitimate users.

Social bots are not always problematic. They are same as other bots in their working but their focus is more on building social relations with the online people e.g. politicians can use social bots to get connected with the public of interests, companies can use them as their customer care agents, an individual can use them to highly influence a user or a group etc. Social bots imitate the human behavior to gain the attention (for example followers, friend requests, replies, likes etc.) from their targets and use this trust network to spread content or promote an agenda or a product [49]. Also, social bots play an important role in multiplayer online games to make the game more entertaining and interesting for the game lovers [44]. Bots can also be used to gain the unfair advantages in the online games. Establishing and creating social bots is not illegal until and unless it causes any disturbance to the normal functionalities of the system. For example, once a social bot found involved in diffusing of malicious content over the network, it becomes spam bot, and the owner will be punished with "imprisonment up to three years or fine up to five lakh Indian rupees or both" under the IT Act, section 43(b), section 66 read with section 43(d), (e), (f).

---

[13] http://vle.du.ac.in/mod/book/print.php?id=9205&chapterid=13341

### 2.5.3 Like Bots

A "Like" is a support of a post, product, business, etc. registered by clicking the button associated with that item. Like-bots are just computer programs controlled mostly by advertising companies, politician or a normal user to like their products, promote some agenda or to like their activities. One of the main jobs of like-bots is used to increase the 'likes' on ads or pages, but sometimes they can be used to send messages as well. In several OSNs, one can buy fake likes for their content [51] from different online vendors[14] and the sellers (usually cyber professionals) make use of multiple numbers of like-bots to like the customer content. The number of likes for a product or a page signifies its success and reputation. People use like-bots for their own benefit which can misdirect the normal users. Obtaining fake likes for a product can lead a customer to believe in the quality of that product and prudently setting pseudo followers to your profile can make you influential. Selling and buying of fake like and followers is nowadays a million dollar business. Online sites like socialbuzzstore[15] provide 1000 Facebook page likes for 25 (USD) and five hundred Facebook followers cost 19 USD.

Very few studies have been carried out for the detection of like bots (or fake likes) so far. In [51] the authors have conducted a comparative study of 'likes' of Facebook pages produced by Facebook ads and several like-farms. The authors created more than a dozen honeypot pages on Facebook and analyzed the produced likes based on users' demography, temporal and social behavior, etc. Like-farms can make the use of like-bots for their businesses, but naive users need to be aware of these fake likes, otherwise, they are likely to get unacceptable results. One more study [10] analyzed a number of Facebook accounts used by some Like-farms and compared their contents (posted on their timeline) with normal user content and found that Facebook accounts owned by like-farms mostly produce likes and comments and often post the same content. Creating like-bots for gaining fake likes and bogus followers deceive the customers and can cause potential customers to lose trust on the organization. Nowadays, it is very hard to find that how many 'likes' for a product or a post are from real users and how many of them are fake which results in misdirection and poses a false impact on the normal users. Therefore, a strict mechanism should be employed to identify the like-bots and shut them off. Deceiving the people by means of fake-likes falls under the crime of online deception and Federal Trade Commission (FTC) [102] has the legal authority to charge fines if a company found involved in such kinds of crime.

### 2.5.4 Influential Bots

Influential bots are automated identities that illegitimately perform discussions on some trending topics on OSNs like Facebook and Twitter in order to influence opinion or to popularize the topic [52]. Influential bots usually generate messages (tweets or posts) either by reposting (or retweeting) the content posted by other users on the same network or create their own synthetic message by an already defined set of rules. Nodes (users) who are connected to the maximum number of nodes in the network are called core nodes and these core nodes play an important role in influencing a topic or an individual. Since the influential nodes have one of their goals to spread the content to the maximum number of people, therefore, they try to send a maximum number of friend/connection requests before spreading the content. Influence of a particular node (user) depends upon its popularity and level of trust in the network and the popularity of a node is determined by the number of incoming requests or received messages [54].

[14]https://boostlikes.com
http://kingdomlikes.com
[15] socialbuzzstore.com

Influential nodes play an important role in viral marketing, but for marketing companies to identify influential nodes on a network often seems to be challenging. Therefore nowadays the organizations first design their OSN bots and start getting into online communities to reach the maximum number of customers. Once these bots obtain trust level within the real users in the network, they start promoting products or brands.

The main job of influential nodes in the network is to change the opinion of users about a particular topic or product. Influential bots, in the same way, try to change the way the people think about an article or any brand on an OSN. Since the normal (real) influential users and the influential bots have almost the same job, therefore it is possible that they have some set of features in common. One possible way to identify the influential bots is to make use of tools and technologies like Klout[16] and Twitalyzer[17] which are used for normal influential identification. Various studies have been carried out for the identification of influential users in online social media like [55], [56], [57]. The more dangerous fact about "influence bots" is that they are being used systematically to influence any online debate by making something appear popular when it is not.

### 2.5.5 Botnet

The network of automated computer programs in an OSN is referred as Botnet. Each program (bot) in this network is assigned either a similar or different set of tasks to be performed in an automated manner. A botnet is a collection of computer programs handled by a 'control-channel' which gives commands to perform unlawful activities [47].

Since the botnet consists of multiple bots, therefore the botnet controller can perform different kind of tasks like spreading malicious content (spam bot), liking a post or a product (like-bot), sending friend request to unknowns (interaction/social bot) and popularizing a topic (influence bot) at the same time.

Botnets are mostly controlled by malevolent users called 'botmasters' by issuing commands to perform malicious activities. Basically, the main purpose of botnets was to assist the users in Internet Relay Chat (IRC) chat rooms [47] by controlling the interactions, providing help to administrators, offering games, extracting information about the platform (operating system), and other details of the user such as email addresses, logins, aliases etc. In [66] the authors have studied the growth of social botnet in the Twitter network and observed how the tweets of a normal user differ from the content generated by a social botnet and how these social botnets help in popularization.

Historically, botnets were primarily used to spread misinformation, propaganda and for many other malicious activities. Several kinds of bots get infiltrated into the target OSN to start a Botnet campaign. The botnet units (bots) help each other by liking the post without verifying it, influencing (retweet or share) the content of each other, writing the positive comment or review etc. in order to gain the trust in the targeted OSN. Botnets are mostly designed for different kinds of benefits varying from individual to individual e.g. shopping companies design them to get likes and increase the ratings of their products, researchers, academicians and data scientists use botnets to crawl data from the web, hackers and other cyber criminals use them as tools for social engineering. A study [53] designed the botnet with three components namely socialbots, botmaster and control-and-command-channel which handles the targeted OSN profiles, providing commands (like posting a message, sending friend/connection request, etc.) and carrying the commands respectively. The botnet has been designed to extract the data from the internet.

A pictorial representation of a botnet is shown in Figure 5. There are three types of users in the diagram viz. normal users, infected users, and bots. Botmaster is simply the user (adversary) who owns and controls the botnet and provides the commands via command and control channel. The commands are followed by each bot in the botnet.

[16] https://klout.com
[17] http://www.twitalyzer.com

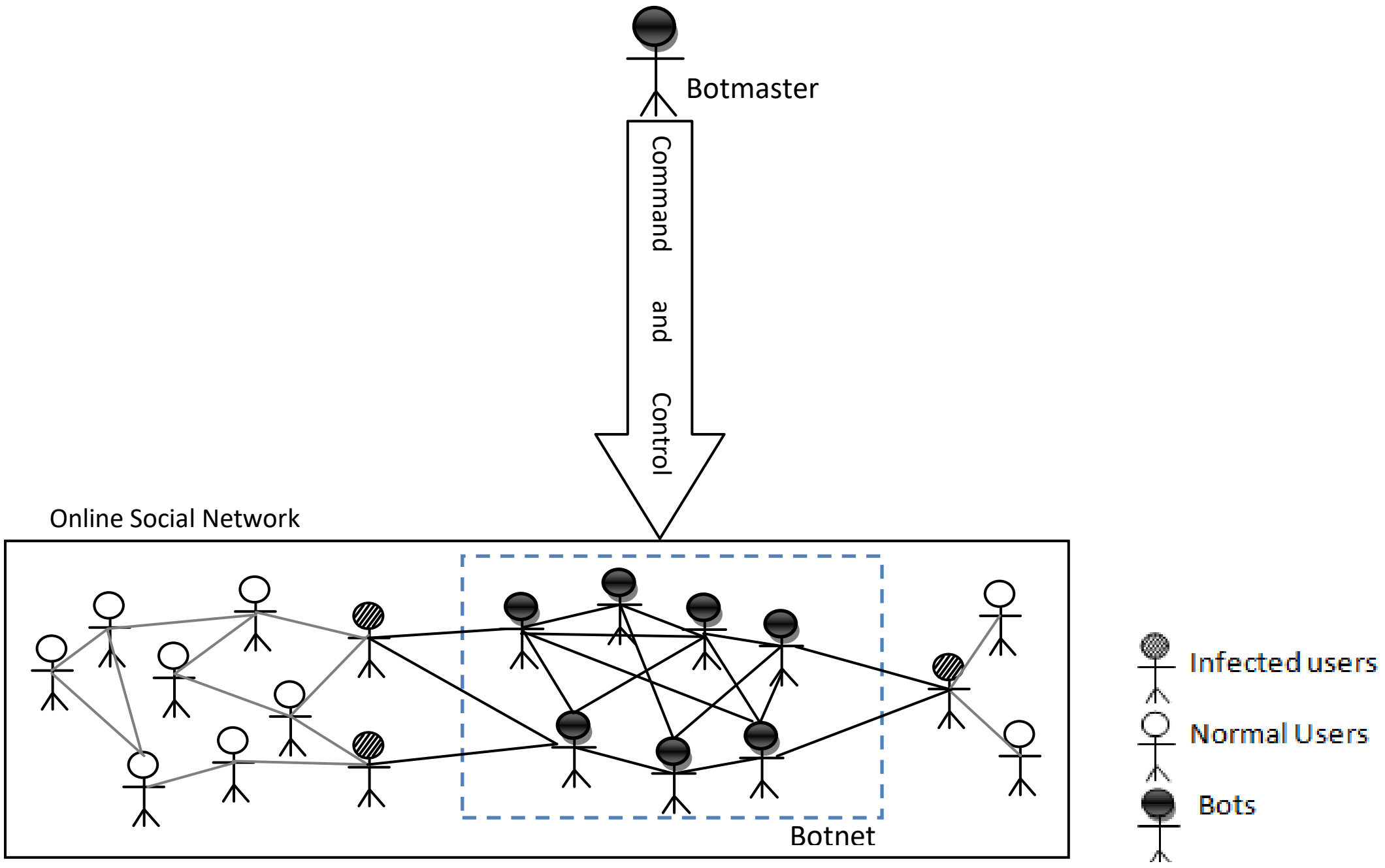

**Figure 5:** Pictorial representation of an OSN Botnet

Bots exploit OSNs as an attractive medium to spread the abusive content, bias public opinions, influence user perception and perform fraudulent activities, etc. and are very complex and highly evolving threats to users' trust and security on the internet. Therefore, serious strategies and steps should be taken to mitigate their effect and risk associated with them. Since OSNs are for real humans, handling profiles by automated programs is against the rules and regulations of OSNs. Bots, either used for commercial activities, entertainment or research must obey the cyber rules and regulations. Table 1 summarizes different kinds of bots and the group of people who mainly use them along with the type of network where they are mostly found.

| | **Social Bots** | **Spam Bots** | **Like-Bots** | **Influential Bots** | **Botnets** |
|---|---|---|---|---|---|
| **Purpose** | To create social, personal and professional relations | To spread malicious content | To increase the ranking/ratings<br>To gain false followers | To alter the behavior of people<br>To perform viral marketing | To perform various unlawful operations by bot network |
| **Used by** | Politicians,<br>Researchers<br>Academicians | Cyber criminals,<br>Marketing and advertising agencies. | Marketing and advertising agencies | Politician<br>Marketing and advertising agencies,<br>Celebrities | Researchers,<br>Hackers,<br>Marketing and advertising agencies, etc. |
| **Networks** | OSNs, Discussion forums. | OSNs. | e-commerce sites, OSNs. | OSNs, Discussion-forums, e-commerce. | Discussion forums, OSNs. |

| References | [19, 27, 44-47, 49, 50, 52] | [50,48] | [51,52] | [50, 36, 34, 52] | [10, 47,50, 52, 53, 66] |
|---|---|---|---|---|---|

**Table 1:** Fake Bots and their characteristics

Fake profiles have been seen very risky for both the OSN service providers as well as their users [33], [34] and can be more dangerous in future if not detected at early stage. As soon as one creates an OSN account, he/she becomes susceptible for targets of an adversary. The fake profiles can catch one's behavior and convince the user to perform unlawful activities. From above discussion, it can be concluded that fake profiles are basically of two types; one created manually and the others using automated methods. And automatic fake profiles pose more threats than other kinds of fake profiles. A botmaster can handle several fake profiles simultaneously (botnet) which damages the reputation of the network to great extent. Therefore, in order to assure the privacy and security of user data and reputation of the network, it is suggested to focus on characterization and identification of automated fake profiles.

So far we have seen five different types of fake profiles and their different characteristics exploited by adversaries in order to perform illegitimate activities. Table 2 summarizes these malignant profiles and their main goals on the network along with the group of people/organization that are affected by them.

| | **Compromised Profiles** | **Cloned Profiles** | **Sock Puppets** | **Sybil Accounts** | **Fake Bot Profiles** |
|---|---|---|---|---|---|
| **Definition** | Existing legitimate profile taken over by an adversary. | Duplicate profile of existing, legitimate profile created by cloner. | Fake account developed with an intention to deceive others | Multiple forged accounts manually established and controlled by a malicious user | Software program designed to control the fake profile to perform malicious activities automatically. |
| **Purpose** | To defame or steal personal information of a person.<br><br>To spread malicious content by exploiting the trusted network. | To defame or steal personal information of a person.<br><br>Fun and Entertainment | To honor, defend or support a person or an organization<br><br>Manipulate a Public opinion.<br><br>To avoid a suspension or ban from a website. | Bad mouthing an Opinion<br><br>Casting fake votes<br><br>To spread malicious content. | To perform viral marketing (influential-bots).<br><br>To increase the number of fake likes. |
| **Types** | Partial-Compromised(PC), Complete-Compromised(CC) | Intra-site cloning, Inter-site cloning | Strawman-sockpuppet, Meatpuppet | -- | Spam-bots, Social-bots, Like-bots and Influential-bots |
| **References** | [7] [58] | [22][24][25][28][29] | [2] [39][40][41][67] | [36][37][38][42][68][77] | [10, 44-53, 99 ] |

**Table2:** Summarization of various fake OSN accounts

In the cyber world, it has been seen very challenging to recognize any activity as a pure crime. Although it is not possible to completely eliminate cybercrime, with the evolution of cyber space, new laws should be incorporated and embedded into existing ones to make them stronger. Besides, the existing cyber laws are associated with several shortcomings such as, there is ambiguity in terms, clear definitions for the important terms in law are not mentioned which can be dangerous and may have several degrees of interpretation. For example, online libel does not define which situations are considered as libelous. Also, the law by creating fear, that any negative commentary and criticism may be seen as an "attack," and inhibits those persons who want to speak out. Sometimes, depending upon the way it is treated, the exposition of the truth may be seen as libelous.

Table 3: Fake profiles and Cyber Crimes along with Cyber Laws

| **Fake Profiles/ Bots** | **Cyber Crime** | **Cyber Law(s)** |
|---|---|---|
| Compromised Accounts | Phishing, Defamation, Spamming. | Section 66D of the Information Technology Act, 2000<br>Section 449 and 500, Indian Panel Code (IPC)[18] |
| Cloned Profiles | Identity theft, Spamming, Defamation, Stalking | Section 66 C, Section 455, information Technology Act, 2000,<br>section 449 and 500, Indian Panel Code (IPC), |
| Sybil Accounts | Astroturfing, Cheating, Fake followers, fake reviews, Deception. | section 415[19], section 417[20] Indian panel Code (IPC) |
| Spam Bots | Spamming, pornography, Astroturfing, malware | Section 449 and 500, Indian Panel Code (IPC),<br>CAN-SPAM Act of 2003 [98],<br>Section 43(b), section 66 read with section 43(d), (e), (f). |
| Like-Bots | Deception, Fake promotion, Astroturfing | Section 415, Section 449 Indian Panel Code 1860 |
| Sockpuppets | Astroturfing, Deception, etc. | Section 5, Federal Trade Commission (FTC) [102] |

As we already discussed in the paper, the punishment for a crime depends upon its intensity, level of damage to the system or the society. Various laws have been outlined for different cyber crimes. The Table 3 summarizes different kinds of cyber crime performed with the help of different kinds of fake profiles along with the cyber laws to curb them.

## 3. Fake Profile Attributes

For detecting fake profiles on social networking websites, the prime requirement is to analyze the characteristics which will distinguish these profiles from the real ones. In order to design an efficient fake profile detector, one must prepare an appropriate and effective feature set. The features can be either observed manually from the social network sites or explored using literature survey. However, it is also possible that some of the features existing in literature may not prove to be fruitful at present as the adversaries are keep on changing their behavior to fool and bypass detection systems. Different features of online profiles have been identified by several researchers from time to time to train their fake profile detection models [2, 8, 31, 32, 35, 39, 40, 41, 44] and based on their nature, we have categorized them into 5 classes as discussed in following subsections.

[18] http://bombayhighcourt.nic.in/libweb/oldlegislation/ipc1860/Chapter%2021.pdf
[19] http://thepracticeoflawjalan.blogspot.in/2012/04/offences-cheating.html
[20] http://devgan.in/ipc/section/417/

### A) Network Based Attributes

As in real life, people interact with friends, make new friends, discuss issues, etc., the OSN users also perform their daily social activities like interacting with online friends, making new friends, and joining new communities (groups, pages, etc.) and creating a network of trust within friends. The features which represent the structural aspects of a social network are called network attributes, for example, groups joined by users, number of friend requests accepted (in degree), number of friend requests sent [8] (out-degree), the extent to which a node acts as a bridge between other nodes (betweenness centrality), nearest node to all other nodes in the network (closeness centrality), etc. Based on these network attributes various researchers designed models for the detection of fake profiles on OSNs. Network features are recommended for the identification of fake profiles in Twitter network. A study [5] towards the characterization of real profiles have identified three features including the growth of OSN friends with time, authentic social interactions and change in the structure of OSN graph over time. Under the third feature, the average degree of nodes and number of singleton friends are taken into consideration to detect the fake profiles. Different users have a different set of features, depending upon the type of social network they belong to. Therefore, to spot out the malicious users in a specific social network, a study of the environment and the culture of that particular network would be more helpful.

### B) Content Based attributes

What user posts or shares on the profile such as text, photos, videos, etc. is referred as content, for example, numbers of tags in the post, the number of words in the post, etc. The content tells a lot about the behavior of the user. Not only the behavior but also the way of thinking, even overall personality of a user on the network is reflected by his/her content. In the literature, researchers have employed several content-based attributes to identify various kinds of spammers on different social networks. In [41], the authors have used content-based features like use of capital or lowercase, quotation-count and punctuation mark for the detection of sockpuppet accounts in Wikipedia. Another study [8] on a Chinese social network Sina Weibo used attributes like the number of hash tags and URLs in the post in order to detect the spammers on the network. Similarly, the content-based attributes like a number of photos of a person tagged in, the number of tags in the uploaded photos by the person have been used in [21] for the detection of fake profiles in the Facebook social network. As per the study of authors in [59], if any account posted a duplicate content, it can be considered as spam account because genuine accounts do not usually update same content multiple times. Another feature can be the message title. Twitter has the “hash tag” mechanism by which we can say what a particular post is about. Usually, the users discuss about a particular set of topics of their interests such as favorite sports, favorite movies or some political party, etc. Since the users typically post about their favorite topics and thus are mostly related to each other but on the other hand, fake user-generated posts are usually unrelated. This unusual behavior can also help to spot an anomalous user. The authors in [3] have used content-based attributes such as the ratio of messages containing URL to the total number of messages and similarity among the messages sent by the user to identify spammers on Facebook, MySpace and Twitter networks.

Initially, researchers were more focused towards the URL centric features to detect human controlled fake profiles. But with the increasing number of automated spam accounts, other forms of unlawful content such as malicious text, pornographic pictures, etc. have also been seen frequent on the network. Therefore, several opinion oriented (content-based) features have been leveraged by various researchers for the identification of spammers on the social networks. For example, POS (Part Of Speech)-tags feature is a content-based attribute which assists in identifying adjectives, adverbs, verbs or nouns used as opinion phrases in the text content. The authors in [35] have used a feature like pages liked by the user and other features for the detection of anomalous user behavior on the Facebook network and stated that if all the pages liked by a user strictly belong to a single category only, the user may be considered as suspicious.

### C) Temporal Features

Temporal features as the name implies are the characteristics related to time, for instance, account creation time, last login time, the time between the two status updates, active account time, etc. According to authors in [58], the users who post their updates or showing any other activity on their profile in odd periods (regular sleeping hours) can be considered as suspicious. Also, in case of social bots, it has been seen that a group of accounts (botnet) is getting active at the same time, perform activities (usually malicious activities), and logout at the same time because social bots are controlled by a single adversary. This time-based behavior can help researchers to identify a social botnet.

### D) Profile Based features

Profile features refer to the basic information about the user identity on an OSN such as gender, location, age, phone number, email address, nationality, profile name, profile picture, number of friends, work and education, etc. Several studies [12, 21, 44] have used user profile attributes to distinguish between normal and anomalous users on different social networks. In [96], the authors have applied several machine learning techniques on profile-based features, friend information (such as the number of friends, number of followers, etc.) as one of the predictor variables to expose spammers on MySpace and Twitter network. In [44], a machine learning based approach has been used for the detection of social bots in the Twitter network where the author has used profile feature like the ratio of followers to the followee of a user profile. Furthermore according to Twitter spam policy, if the number of people following you is less than the number of people followed by you, or you trying to follow the people beyond the limit [3], your identity can be considered as suspicious. Similarly, in a study [12], the authors used profile features such as following/follower ratio and account age along with other types of attributes to analyze distinguishing characters of spammers on two Chinese microblogging networks, Sina Weibo (weibo.com), and Tencent Weibo (t.QQ.com). Apart from above list of features, literature is replete with other profile characteristics which can be used to identify anomalies in several OSNs, for example, languages known by the user as in certain scenarios, some users mention multiple languages in their *"languages-known"* column, later they write posts in some other language, which is not considered as the normal behavior.

### E) Action Based Features

Action-based attributes refer to the activities performed by the user on a social network, for example, posting, uploading or sharing, etc. Reacting or commenting on friend's posts also reflects the actions of a user on the network. Action-based features also called activity features [30] play a fundamental role in identifying spam accounts as they perform unlawful and discriminatory activities publicly. Another study [8] on a Chinese social network Sina Weibo uses attributes like number of messages per day, the number of comments, number of likes in the post, etc. to detect the spammers on the network. It has been observed by the authors of the study that the spammers post messages three times more than the non spammers. In [35], the authors have used the behavioral feature like how frequently the user likes the pages (rate of like activity) for the detection of anomalous user behavior on the Facebook network. If the frequency of page likes within a specific time interval is very high, the user may be considered as suspicious.

All the 5 categories of attributes discussed above have been summarised in Table 4 below along with the category of fake profile that has been detected by using those features on different social networks. This exhaustive list of identified features would certainly be helpful for researchers to build different fake identity detection models for social networks.

**Table 4:** Different categories of features and type of fake profile identified by them

| **Features/Attributes** | Author(s) | Network | Targeted fake profile Category |
|---|---|---|---|
| - Profile layout colors (P)<br>- First names, user names (P)<br>- Spatiotemporal Information (T) | [31] | Twitter | Spam bots |
| - Pages liked by the user (C)<br>- Rate of like activity (A) | [35] | Facebook | Sybil Accounts |
| -Writing style (C) | [39] | Tianya<br>(China forum) | Sockpuppets |
| - Number of Replies (A)<br>- Registration Dates (P) | [40] | Uwants<br>(Hong Kong discussion forum) | Sockpuppets |
| - Ratio of friend requests sent to the number of Friends (N)<br>- Ratio of messages containing URL to the total number of messages (C)<br>- Similarity among the messages sent by the user (C)<br>- Number of messages sent (A)<br>- Friend Count (P) | [3] | Facebook<br>Twitter<br>MySpace | Spam bots |
| - Number of total revisions (A)<br>- Article discussion (C)<br>- User page (P )<br>- User discussion page (C) | [2] | Wikipedia | Sockpuppets |
| - Punctuation count (C)<br>- Quotation count (C)<br>- Use of capital or lowercase (C) | [41] | Wikipedia | Sockpuppets |
| - Number of friends (P)<br>- Number of followers (P)<br>- Follower ratio (P) | [44] | Twitter | Spam bots |
| - Number of reposts (C)<br>- Number of Comments (A)<br>- Number of Likes (A)<br>- Number of Mentions (C)<br>- Number of URLs in the post (C)<br>- Number of Hashtags (C) | [8] | Sina Weibo | Sockpuppets |
| - Education and work (P)<br>- Relationship status (P)<br>- Gender (P)<br>- Number of wall posts by the person (A)<br>- Number of photos of person tagged in (C)<br>- Number of photos the person has uploaded (A)<br>- Number of tags in the uploaded photos by the<br>- Person (C) | [21] | Facebook | Sybil Accounts |
| - Number of Followers (P)<br>- Number of Followees (P)<br>- Number of Friends (P)<br>- Number of microblogs to get a fan (A) | [12] | Sina Weibo and Tencent Weibo | Sockpuppets |
| A: Action-based Attribute, C: Content-based Attribute, N: Network-based Attribute, P: Profile-based Attribute, T: Temporal Attribute | | | |

## 4. Data Collection Approaches

So for we have seen different types of features used in research particularly for the detection of fake suspicious identities in OSNs Collecting a required (for example, specific to fake profiles) dataset is considered as a prime challenge while analyzing Online Social Networking sites. Different techniques have been employed by researchers to extract the data from ONS sites. A study [113] has presented a detailed discussion on web data extraction techniques along with the application domains in which they are applied.

In this section we discuss different approaches to collect the required data of both fake as well as real profiles from a social network. The most popular methods to collect the required data includes data extraction using APIs (Application Programming Interfaces) provided by the service providers, designing standalone crawler program, generating artificial data using available tools, or using existing datasets as shown in Figure 7. All the four methods are briefly explained as under:

### 4.1 Data Collection using APIs

Collecting data using APIs is mostly used nowadays for social network analysis and is highly recommended. Generally, the OSN service providers assist developers and normal users with the several libraries (packages) for various data extracting operations. In order to collect the problem specific data from a social network, most of the researchers write their own code to interact with the targeted social network via API. Almost every social networking site has its own API, for example, there is GRAPH API[21] for the Facebook network which allows its users to interact with their application and collect user information. Similarly, for Twitter, there is Twitter REST-API[22]. The working of an API is like a web site, we can pass the requests to it and get back a response over the HyperText Transfer Protocol (HTTP).

In [60], the authors have used Twitter API methods to crawl and collect several activities of users such as 100 most recent tweets, following, followers, etc. A similar approach was used in [5] in which authors developed a Facebook sensing application to collect the required statistical information from user's Facebook wall via Facebook Graph API. The researcher in [61] has designed and implemented an application called "NetVizz" based on the Graph API for the collection of information about Facebook users.

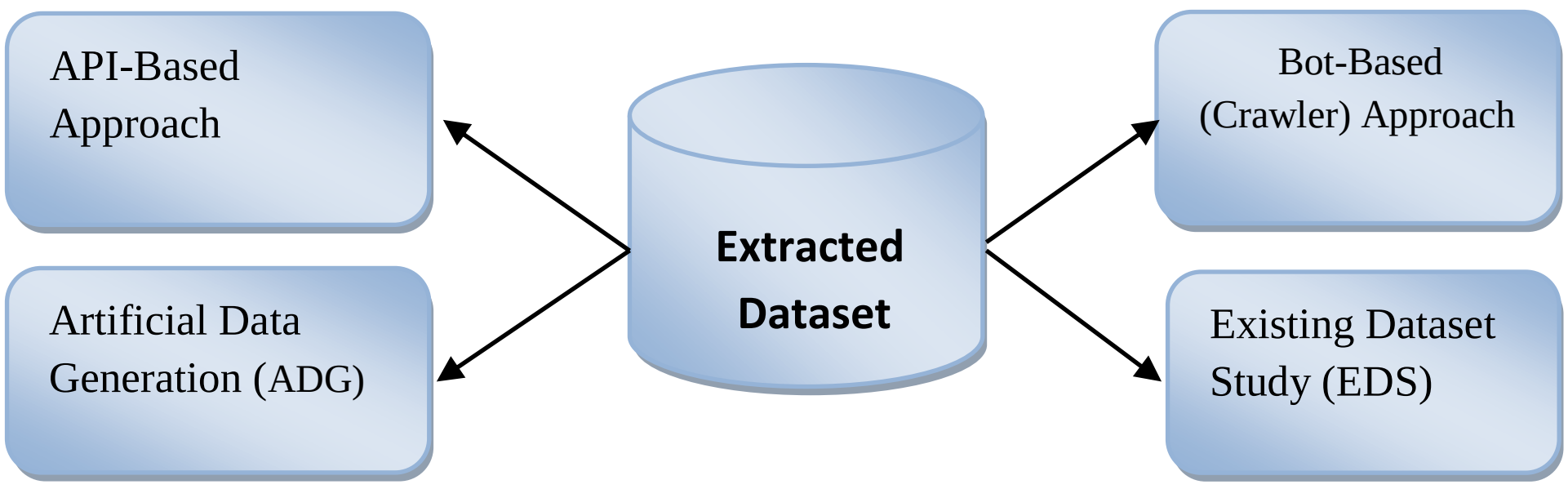

**Figure 7**: Data Collection Techniques

[21] https://developers.facebook.com/docs/graph-api
[22] https://dev.twitter.com/rest/public

No doubt APIs are fast and easy to use but unfortunately, these APIs are associated with several unavoidable constraints such as data request rate restrictions, selective data access, etc. The limit on request rate may inhibit the user to obtain dataset timely. Also, every piece of information on the network may not be retrieved via these APIs. Therefore, it is not always necessary that one will get the required dataset by using APIs provided by OSNs. In that case, there are other alternative approaches which are discussed in the following sections. Figure 8 (a) gives a conceptual view of API-Based data collection approach for social networks.

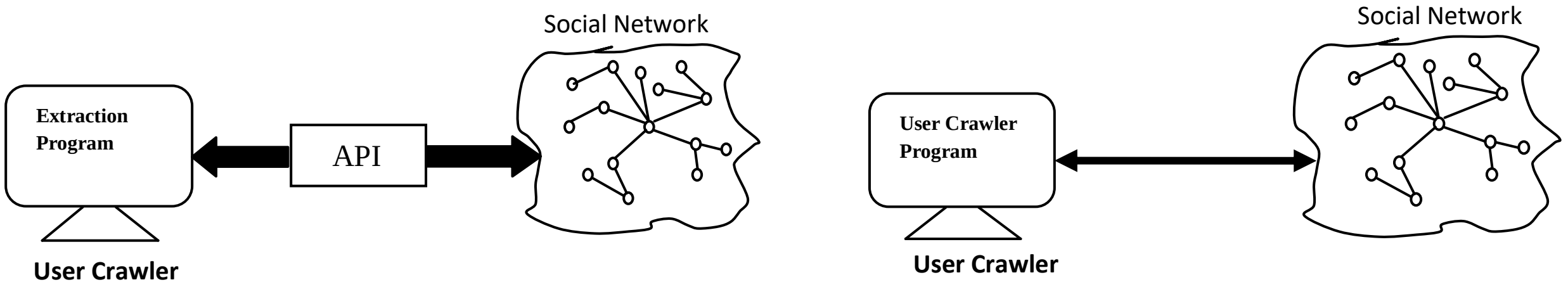

**Figure 8**
**(a)** Data extraction using API **(b)** A typical OSN data crawler

## 4.2 Bot-Based (Crawler) Approach

The bot-based approach involves designing of a standalone data crawler to extract the information from the social network. Like API-based approach, it also extracts the information about users but here the crawler program does not use any API to interact with the social network, rather the direct communication takes place between crawler program and the social network as shown in Figure 8 (b).
Different programming languages may be used to design the data extraction programs such as javascript, python, PHP, etc. However, every extraction program requires a set of seed profiles which are usually selected based on some criteria such as a large number of friends, location-based profiles, etc. The seed profiles are used by the program in order to traverse the network and extract the information. The network can be explored by several graph traversing techniques including Breadth First Search (BFS) [19, 61], Depth First Search (DFS) [109], Forest Fire (FF) [97], etc. BFS technique is one of the most popular techniques employed by researchers for crawling social networks. It starts at the target profile (seed node) and explores its neighbor nodes first, before moving to the next level of neighbors. In case of DFS crawler, instead of extracting all the neighbors of the target profile, the attributes of neighbor's of the neighbor profile up to a specific level are collected first.

The extraction program or generally, we can say a data crawler needs three things. First, a source file that contains the URL of target profiles (seed profiles). Second, it requires data fields to be extracted from a user profiles. And third, a file to store the extracted data. The conceptual view of data extraction program is shown in the figure as under. The crawler program generally makes use of Document Object Model (DOM) for determining the information t m a user profile.

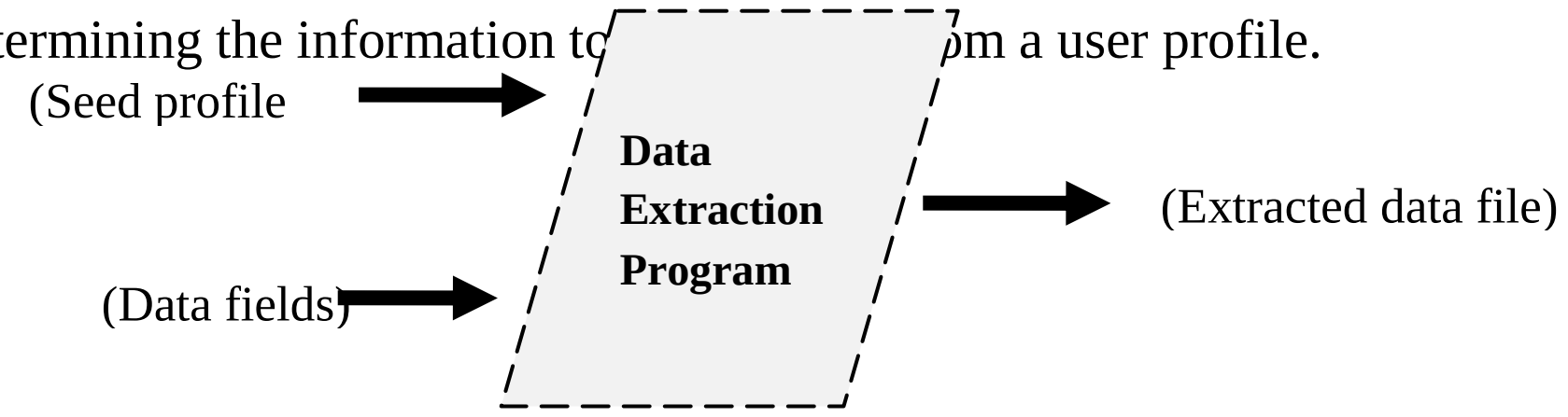

Figure 9: Pictorial representation of Data extraction program

DOM represents a web page in the form of a tree where HTML tags are denoted by nodes and tree hierarchy represents the alignment of the data elements. In [3], the authors have written a script to get connected with already created honey profiles and extracted all the information needed to detect the malicious activities. The authors in [14] have used java API "HTML Parser" to collect the users' public information from the Facebook network. The authors in [115] have designed an iMacros–based data crawler called *IMcrawle*r to extract the information of users on the Facebook website in order to perform the user behavioural analysis on the network.

Although, the Bot-based approach overcomes the limitations of API-based approach there are several inherent shortcomings with this approach as well. First, writing a complete stand-alone crawler program is a very complicated task as one need to start it from scratch. Second, most of the social networks have imposed a restriction on automatic data retrieval, for instance, Facebook has implemented a Facebook immune system to notice any automated activity on the network. Third, the data collected by a bot-based approach is generally unstructured in nature which demands strict data pre-processing.

### 4.3 Artificial Data Generation

API-based and bot-based approaches are time-consuming methods for data collection and are highly subject to privacy and security settings of users. As in many cases, we need data instantly to solve a particular problem, but the data cannot be at hand all the time. Also, one may not have access to the data of interest because of the privacy concerns. Therefore, in such cases, we generate the synthetic data sample based on the structure of a network or the characteristics of existing datasets with the help of existing data generator packages [64,110, 111] or by designing synthetic data generators [93].

The data can be generated using various available tools based on the known statistics or parameters of any existing social network. For example, if we know the degree distribution, clustering coefficient, average betweeness centrality and other statistical parameters, a dummy data set can be generated for analysis purposes. The various online data generators like GEDIS Studio [110], Databene Benerator [111], etc. are available for the generation of artificial data[23]. However, the artificially generated will not be useful till its resemblance is not verified with the original data.

### 4.4 Existing Dataset Study (EDS)

The researchers can also conduct different studies using the data that have been collected by others and made available to the public for the further use. The process is called secondary analysis, where the study is conducted on the dataset generated by others. Authors in [90] used the public dataset to conduct analysis on social bookmarking website-BibSonomy for predicting the spam users. The use of existing datasets saves a considerable amount of time and effort of researchers. Moreover, since the available dataset has already been used for analysis, it contains less noise, outliers, and missing values, etc. and therefore, does not demand rigorous pre-processing. Several researchers have generated the data for their analysis and made it available for others [89, 91]. Also, there are a number of resources on the web that serve as the repository of various publicly available datasets [70], [72], [74], verified and authenticated by research communities for solving various social network analysis problems. Some need a registration process or a request via email [71], [73], [75]. However, the main disadvantage of the existing datasets is that it may not include all the features what an investigator is interested in. Since we are more focused towards fake profiles and the data in existing datasets hardly contain all the features that better serve the purpose, therefore, the existing datasets are least used for fake profile detection in OSNs.

The selection of data collection technique mainly depends upon the data of interest. After the identification of features for the detection of fake profiles, we need to make sure if the data related to these features is existing in any of the available datasets or can be generated artificially. In that case, we directly get our sample data. If neither of the two techniques (artificial data generation and existing datasets) solves the problem, there are two alternative techniques, API-based and bot-based, to deal with the problem. The API-based approach may restrict the researchers to extract every piece of information of their interest while as the bot-based approach enables a user to extract every piece of information visible on a browser but at the expense of technical complexities. However, in both the two approaches data is explicitly scrapped from user profiles by writing an extraction program. Therefore, a set of fake and real profiles need to be selected for the extraction of information. Real user profiles can be easily obtained from our friend list, follower list, trusted people, etc., on the other hand for selecting a set of fake profiles, we have suggested three methods which are discussed in the following subsection. It should be noted that data collection process may involve more than one data collection technique to harvest the data based on the problem under consideration.

[23] http://www.freedatagenerator.com, http://www.generatedata.com, https://www.mockaroo.com

### 4.5 Profile Selection Approaches

It is now clear from above subsections that, existing datasets and artificial data generation methods do not need any kind of profiles to extract the data while as the rest of the two approaches (API-based approach and crawler-based approach) require a list of profiles (seed nodes) to extract data. In order to extract data via API or through a crawler, we need both real and fake profiles. The real profiles are available in the abundant amount on the social network, so can be easily located. One can use the account of trusted users and their network of friends to collect the real profiles. Trusted users can be friends, verified accounts or profiles of personally known users on the network. However, in comparison to real profiles, the fake accounts exist in much lesser amount than the real ones which make it very difficult for the researchers to locate them on a huge network with billions of users. According to a study on Facebook, out of 955 million monthly active users, 8.7% of the users are reported as fake[24]. In order to obtain the list of fake profiles for data extraction, we describe three approaches namely, manual-approach, honey profile-based approach  and botnet-based approach. Honey profile-based approach [3] involves the creation of fake attractive profiles (honeytraps) to attract the people (especially fake) towards them. In case of Botnet approach, a network of interrelated bots (computer programs) is created to interact with other bots and normal users to obtain the fake profile list for data extraction. All the three approaches to obtain fake profiles to be extracted by one of the above-mentioned data extraction approaches are discussed as under.

#### A) Manual Approach

In case of manual approach, we need to explore the suspicious accounts manually and keep a record of the profiles found involved in malicious activities [14]. Generally, in manual selection approach, there are several ways to investigate and select fake profile set. One way is to manually collect a list of random profiles from a network and label every profile in the list based on a set of characteristics which distinguish real profiles from forged ones. For instance, if any user found to be involved in spreading malicious or unlawful content on his/her profile or friends profile, it may be fake and later the information is crawled from it using any of the data collection methods.  In another way, the researchers extract the user attributes from the network and can create fake profile set with the help of domain experts by carefully inspecting each attribute of the collected dataset [94]. The manual fake profiles selection approach also includes those accounts which had already been detected and labelled as fake by OSN service providers or the research organizations[25].

#### B) Honey Profile Based Approach

Honey profiles as the name indicates are the OSN profiles used to attract other (most likely similar) users towards them. Different types of honey profiles or simply honey-traps are established to attract both real and fake users as per the requirement. For example, some people create honey profiles to attract specific group such as teenagers and young people on the targeted network while some people create honey profiles to attract the general public.  However, in case of fake profile selection, the researchers create honey profiles such as porn-based profile which specifically entice the fake ones of the same category [30]. To make the honey profiles active all the time the owners constantly keep on updating these profiles with latest and interesting stories and images. The authors in [3] have created around 900 honey profiles on three different social networks viz Facebook, MySpace and Twitter and then analyzed the collected profiles to identify the anomalous behavior of users who contacted their honey profiles. Similarly, in [9], the authors created eight very interactive and attractive profiles on Facebook with different age groups to collect the user profiles and observed these eight profiles for a period of three months. Another study [51] has used honey profiles to collect and uncover social spammers in social networks. As the honey pots attract users' attention on the large scale, therefore, they are mostly expolited by spammers to spread malicious activities as well.

[24] https://www.digitaltrends.com/mobile/8-7-percent-of-facebook-users-are-fake/
[25] https://www.barracuda.com/

### C) Botnet Based Approach

As already discussed, a botnet is a network of automated programs (bots) controlled and supervised by a human controller called 'botherder', designed to perform a number of tasks such as interacting and attracting other users on the network, promoting products and brands, election campaigning, etc. [53]. Unlike above two approaches, where we select already existing fake profiles on the network, the botnet-based approach involves the injection of a series of automated interacting fake profiles into the network and increasing the trust level among other users on the network to obtain the fake profile list. Unlike honey profiles which most of the time wait passively for the specific connection requests, the profiles in botnet are mostly active and are constantly involved in tasks like sending connection requests, generating content [47], etc. in order to gain large user base. Moreover, the honey profile-based approach is effective to attract a specific domain of profiles such as pornographic fake profiles or other adult content spreading profiles, etc., but would not give satisfactory results in attracting domains like astroturfing accounts. On the other hand, botnet approach can be used to simulate the behavior of broader domains of fake profiles. Later on, the botnet profiles as well as the fake profiles attracted by botnet are used to construct a fake profile dataset. In [53], the authors have designed a botnet with three components, socialbots, botmaster and control-and-command-channel to show the vulnerability of OSNs to an infiltration by socialbots. The pictorial representation of a botnet is shown in Figure 5.

## 5. Fake Profile Detection Techniques

In section 2, several types of OSN profiles and their properties have been described. Here, in this section, we highlight category wise, the different machine learning techniques, employed by researchers for the identification of different categories of fake profiles on online social networks.

It has been observed that compromised accounts play a vital role in spamming as these users exploit the level of trust maintained by the real users on the network. In the paper [58] authors proposed a tool, COMPA to detect compromised accounts on Twitter and Facebook. The authors have used statistical modelling to build the behavioral profile of users on the basis of characteristics of their sent messages and use several similarity metrics like n-gram analysis to compute the anomaly score. The authors have used Sequential Minimal Optimization (SMO) to determine the weights of the features in the dataset. In paper [23] the authors have detected spam campaigns on Facebook by applying clustering techniques on the wall posts. They further analyzed each malicious account for the presence of compromised accounts based on the content (photos, videos, etc.) generated on their walls.

A number of researchers have focused towards to detection of cloned accounts on the social networks. In [25], authors have used Markov Clustering algorithm (MCL) to divide the Facebook network into smaller communities based on their similarities, all the profiles similar to the real profiles are gathered to calculate the strength of the relationship in order to check whether it is a clone or not. In another study [29], authors propose a method for detecting social network profile cloning by designing a system with three components namely *information distiller, profile hunter, and profile verifier*. Information distiller extracts the information from real user profiles and selects attributes which can be used to uniquely identify the profile. Profile hunter processes the information and locates the profiles of the user on different OSNs. Profile verifier calculates the similarity score between all the profiles and presents the result to the user.

Several studies have also been conducted to identify sockpuppet accounts on social networks. A study in [39] has created a network of users based on their topics of interest on Tianya forum and Taobao online auction website. Based on the writing style of the users, the author pruned the graph to obtain sockpuppet network (SPN). Finally, the different community detection techniques have been applied on SPN to identify sockpuppet communities. Another method to detect sockpuppets on a Hong Kong-based discussion forum is presented in [40]. The method is based on the total number of topics posted by one account and the number of replies from the other accounts. A detection score is calculated to spot a sockpuppet pair. The larger the score, the more will be the chances of two accounts being the sockpuppet pair. Based on the verbal features like of the user's Punctuation count, Quotation count, Use of capital or lowercase, the authors in [41] presented a sockpuppet detection method for Wikipedia network using natural language processing techniques.

Researchers also paid a vital attention towards the detection and mitigation of Sybil accounts from social networks [36, 42, 68, 77], but the detection of Sybil attacks is still in its infancy. Most of the Sybil defense techniques work on the ranking of nodes based on how well a node (an account) is connected to trusted nodes (legitimate accounts), the node has a higher rank if it is within the local community of a trusted node. In SybilGuard [38] the authors present a novel approach to protect a social network from Sybil attacks. They consider the link between two nodes as a trust relationship. Sybil nodes are differentiated from trust nodes using the calculated trust-relationship. Basically, the SybilGuard depends upon two characteristics of underlying social networks, first, the trusted accounts always have a huge number of links, second, the fake users create many nodes (accounts) but with few trusted connections. The authors in [30] have used Markov Clustering (MCL) technique on real data of Facebook network for the identification of fake profiles using the features such as Facebook fan pages, links shared and active friends, etc. MCL technique groups the users into three clusters, one contains all the fake identities, second contains all the normal profiles and the third cluster contains the mixture of both. This study suggested that techniques like a decision tree, Support Vector Machine (SVM), Naïve Bayes (NB) etc. can be a choice to classify profiles as fake or real, but do not work efficiently for the social network profile dataset with multiple classes. Also, there is a scarcity of such well-defined profile data sets and most of the data set neither have any predefined class label nor a well-defined feature set, therefore unsupervised learning techniques are preferable over supervised techniques. Similarly, a novel approach has been proposed by authors in [31] for deception detection in a Twitter using gender and location attributes by applying Bayesian classifier and k-means clustering. Authors in [3] have designed a classifier based on Random Forest (RF) algorithms to identify the spammers among users who got connected with their honey profiles. Based on the spam strategy, four categories of spam profiles were distinguished namely Displayer, Bragger, Poster, and Whisperer.

Spam bot detection is also taken into consideration by several researchers. The authors in [44] have shown that out of four (decision tree, support vector machine, K-nearest neighbor and neural networks) classification techniques, the Bayesian classifier is the best in predicting spam bots in Twitter network. In [10] the authors have studied the growth of social botnet in the Twitter network and observed how the tweets of a normal user differ from the content generated by a social botnet and how these social botnets help in popularization. The authors of a study [52] proposed a framework that comprises of three steps for detecting influential bots on a social network. In the first step, the bots are identified by basic techniques like manual inspection, behavioral analysis or linguistic knowledge. In the second steps, more bots are identified on the basis of clustering, outliers and network analysis. In the third step, remaining bots are identified by building the classifier using the bots and humans identified in the earlier steps as training data.

| **Network** | **Fake Profile Category** | **Technique(s)** | **Author(s)** |
|---|---|---|---|
| Twitter | Spam bots | Bayesian classification<br>K-means clustering | [31] |
| Facebook | Compromised Accounts<br>Sybil Accounts | k-Nearest Neighbor (kNN)<br>Principal Component Analysis (PCA) | [35] |
| Tianya<br>(China forum) | Sockpuppets | Authorship-identification techniques<br>Link Analysis | [39] |
| Twitter | Spam Accounts | Random Forest<br>Sequential Minimal Optimization (SMO)<br>Naïve Bayesian<br>K-NN neighbor | [60] |
| Facebook<br>Twitter<br>MySpace | Spam bots | Random Forest algorithm | [3] |
| Twitter | Spam bots | Random Forests<br>AdaBoost,<br>Logistic Regression<br>Decision Tree | [99] |
| Wikipedia | Sockpuppets | Support Vector Machine (SVM)<br>Random Forest (RF)<br>Adaptive Boosting (ADA) | [2] |
| Wikipedia | Sockpuppets | Support Vector Machine (SVM) | [41] |
| Twitter | Spam bots | Bayesian classification | [44] |
| Sina Weibo | Sockpuppets | Support Vector Machine (SVM) | [8] |
| Twitter | Sybil Accounts | Support Vector Machine (SVM) | [36] |
| Facebook | Sybil Accounts | K-means Clustering<br>Naïve Bayesian Classifier<br>Support Vector Machine (SVM) | [21] |
| Sina Weibo<br>Tencent Weibo | Sockpuppets | Decision Tree-C4.5 (J48)<br>IBk,<br>Naive-Bayes<br>Bagging<br>AdaBoostM1<br>Rotation Forest<br>OneR | [13] |
| Twitter | Spam Accounts | Bayesian classification Algorithm | [59] |
| Twitter<br>Facebook | Compromised accounts | N-gram Analysis<br>Sequential Minimal Optimization (SMO) | [58] |

**Table 5:** Commonly used machine learning techniques for detection of suspicious accounts on OSNs

Apart from above discussed approaches, we have presented few most popular and commonly used machine learning techniques in Table 5, employed in research to detect the fake profiles on social networking websites.

It can be clearly seen from the above table that SVM, Decision tree and Bayesian classification algorithms have been mostly used in the fake profile detection studies. Furthermore, most of the researchers have used Twitter social network to conduct their study. The reason can be the availability of data as the Twitter allows access to the user data for research purpose [48]. Keeping that in mind, this paper also provides different approaches to obtain data from social networking websites and assist researchers to carry out their studies on different OSNs in different dimensions.
Different machine learning based techniques have been applied for the prediction and identification of forged accounts on social networks, but most of the researchers concentrate only on a specific type of fake profile which lefts the other fake profile categories undetected and enables the attacker to contaminate the network. Therefore, a generalized method which can spot maximum types of fake identities on a social network is highly required. Here in this paper, put everything about several kinds of fake profiles at a single place which assists researchers in designing a generalized and an efficient fake profile detection model. Furthermore, with the passage of time, the number of users and the content generated by them on these social networks is growing rapidly and the prediction becomes more challenging. Therefore, identifying the anomalous behavior or conducting any kind of analysis on social networks, highly scalable machine learning techniques for large-scale graph analysis, graph partitioning, and clustering algorithms are needed.

## 6. Conclusions

With thousands of fake profiles on different OSNs having multifaceted aims to deceive, one need to adapt more advanced methods to secure one's online presence as least can be done when the security gets compromised. In this paper, we have described various types of OSN threat generators (fake profiles) like compromised profiles, cloned profiles and online bots (spam-bots, social-bots, like-bots and influential-bots). An exhaustive effort has been taken to put all kinds of malicious entities on OSNs at one place along with existing cyber laws to curb imposters. Since there are very strict regulations and punishments for different categories of cyber criminals but still cybercrime and cyber terrorism dictates across the world. The most important reason for failure in nabbing the cyber criminals is, the investigators are not able to get a trace of the criminals as the crime is mostly conducted across the boundaries of the nation. Therefore, the need of the hour is a worldwide uniform cyber law to combat cyber crimes. This paper also provides a brief outlining of pro and cons of several existing cyber laws which are framed to curb the online fake profiles. Also, to alleviate the data crunch faced by OSN researchers, the paper also highlights different data crawling approaches along with some existing data sources. Furthermore a rigorous survey of techniques used in studies for fake profile detection, has been presented.

Many researchers have tried to mitigate fake profiles to some extent but more concrete steps are still to be taken. It can be concluded that the need for more advanced automated methods still remains unfulfilled for secure social networks. The appropriate and timely steps are needed to develop automated mechanisms to identify suspicious users.